\documentclass[conference,10pt]{IEEEtran}

\usepackage{epsfig}
\usepackage[english]{babel}
\usepackage[utf8x]{inputenc}
\usepackage[T1]{fontenc}

\usepackage{cite}
\usepackage{textcomp}
\usepackage{xcolor}

\usepackage{amsmath}
\usepackage{amsthm}
\usepackage{amsfonts}
\usepackage{mathrsfs}
\usepackage{graphicx}
\usepackage{booktabs}
\usepackage[labelfont=bf]{caption}
\usepackage{subcaption}
\usepackage{verbatim} 
\usepackage{float} 
\usepackage[algoruled,noline,longend,linesnumbered]{algorithm2e}
\DeclareMathOperator*{\E}{\mathbb{E}}

\DeclareMathOperator*{\Sspace}{\mathcal{S}}
\DeclareMathOperator*{\A}{\mathcal{A}}
\DeclareMathOperator*{\Nagents}{\mathcal{N}}

\def\BibTeX{{\rm B\kern-.05em{\sc i\kern-.025em b}\kern-.08em
    T\kern-.1667em\lower.7ex\hbox{E}\kern-.125emX}}

\newcommand{\linebreakand}{
  \end{@IEEEauthorhalign}
  \hfill\mbox{}\par
  \mbox{}\hfill \begin{@IEEEauthorhalign}
}
    
\usepackage[colorlinks = true,
            linkcolor = black,
            urlcolor  = black,
            citecolor = black,
            anchorcolor = blue]{hyperref}    

\graphicspath{{./}{../}}

\usepackage{tikz}
\usetikzlibrary{calc} 
\DeclareRobustCommand{\circled}[1]{%
  {\tikz[baseline=(char.base)]{\node[shape=circle,draw,inner sep=0.6pt] (char) {\fontsize{10pt}{10pt}\selectfont #1};}}
}

\IEEEoverridecommandlockouts

\begin{document}

\title{Is Machine Learning Ready for \\Traffic Engineering Optimization?
\thanks{This work was supported by the Spanish MINECO under contract TEC2017-90034-C2-1-R (ALLIANCE), the Catalan Institution for Research and Advanced Studies (ICREA) and the Secretariat for Universities and Research of the Ministry of Business and Knowledge of the Government of Catalonia as well as the European Social Fund.}
}
\label{title}

\author{
\IEEEauthorblockN{Guillermo Bernárdez\IEEEauthorrefmark{1}, José Suárez-Varela\IEEEauthorrefmark{1}, Albert López\IEEEauthorrefmark{1}, Bo Wu\IEEEauthorrefmark{2}, Shihan Xiao\IEEEauthorrefmark{2}, Xiangle Cheng\IEEEauthorrefmark{2}, \\ Pere Barlet-Ros\IEEEauthorrefmark{1} and Albert Cabellos-Aparicio\IEEEauthorrefmark{1}}

\IEEEauthorblockA{
\IEEEauthorblockA{\IEEEauthorrefmark{1} \textit{Barcelona Neural Networking Center}, 
Universitat Politècnica de Catalunya, 
Barcelona, Spain \\
\{gbernard, jsuarezv, alopez, pbarlet, acabello\}@ac.upc.edu}
\IEEEauthorrefmark{2}\textit{Network Technology Lab.},
Huawei Technologies Co., Ltd., 
Beijing, China\\
\{wubo.net, xiaoshihan, chengxiangle1\}@huawei.com} \\
\textbf{NOTE:} Accepted as a main conference paper at IEEE ICNP 2021. $\copyright$2021 IEEE. Personal use of this material is permitted. \\ Permission from IEEE must be obtained for all other uses, in any current or future media, including reprinting/republishing \\this material for advertising or promotional purposes, creating new collective works, for resale or redistribution to servers\\ or lists, or reuse of any copyrighted component of this work in other works.
}

\maketitle

\begin{abstract}
\addcontentsline{toc}{section}{Abstract}
Traffic Engineering (TE) is a basic building block of the Internet. In this paper, we analyze whether modern Machine Learning (ML) methods are ready to be used for TE optimization. We address this open question through a comparative analysis between the state of the art in ML and the state of the art in TE. To this end, we first present a novel distributed system for TE that leverages the latest advancements in ML. Our system implements a novel architecture that combines Multi-Agent Reinforcement Learning (MARL) and Graph Neural Networks (GNN) to minimize network congestion. In our evaluation, we compare our MARL+GNN system with DEFO, a network optimizer based on Constraint Programming that represents the state of the art in TE. Our experimental results show that the proposed MARL+GNN solution achieves equivalent performance to DEFO in a wide variety of network scenarios including three real-world network topologies. At the same time, we show that MARL+GNN can achieve significant reductions in execution time (from the scale of minutes with DEFO to a few seconds with our solution).

\end{abstract}


\begin{figure*}[h]
\centering
    \includegraphics[width=2\columnwidth]{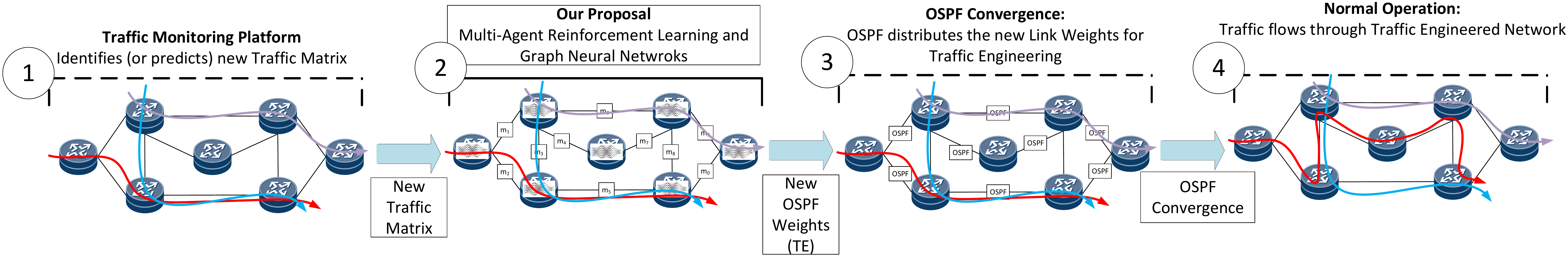}
  \caption{Network Traffic Engineering scenario.}
  \label{fig:implementation}
\end{figure*}

\section{Introduction}
Traffic Engineering (TE) is a well-established mechanism that plays a fundamental role in the performance of today's \mbox{Internet}~\cite{RFC3272}. Particularly, its main goal is to provide efficient and reliable network operations, while optimizing the network resources~\cite{RFC2702}. As a result, there exists a rich body of proposals based on different technologies (e.g., flow-based routing, link-state protocols, overlay networking) that target various optimization goals and network scenarios~\cite{wang2008overview,mendiola2016survey}. Beyond this broad definition, a fundamental TE problem traditionally addressed in the literature is intradomain TE, where the classic optimization goal is to minimize the maximum link load within a self-administered network domain (e.g., a carrier-grade network)~\cite{azar2004optimal, fortz2004increasing,hartert2015declarative}. This is a well-known NP-hard problem~\cite{freeman1991centrality}.
 
The last few years have seen an increasing interest in the application of Machine Learning (ML) to complex network control and management problems \cite{wang2017machine}. Particularly, the outstanding results of Deep Reinforcement Learning (DRL) in other domains (see~\cite{schrittwieser2020mastering} and references therein) have awakened the interest of the networking community in understanding the true potential of this new technology for online network optimization tasks, such as TE (e.g.,~\cite{valadarsky2017learning,xu2018experience,drl-giorgio,geng2020multi}).

In this paper, we raise an open question: {\em Is ML ready for Traffic Engineering optimization?} Here we refer to {\em ready} as achieving -at least- comparable performance and speed to state-of-the-art TE solutions based on classical optimization methods. In order to answer this question, we seek to create a TE solution leveraging the latest advancements in the ML field, and then experimentally compare it to the best TE approach available in the literature.

We present a novel TE optimizer based on a combination of Multi-Agent Reinforcement Learning (MARL)~\cite{foerster2018deep} and Graph Neural Networks (GNN)~\cite{scarselli2008graph}, which can be considered as the most advanced ML technologies for online optimization problems over graphs, such as TE. The proposed solution is distributed over the network devices and is tasked to address the intradomain TE problem. In particular, given a set of estimated traffic demands, the distributed agents of our MARL system cooperate to jointly optimize the link weights used by OSPF~\cite{RFC2328}, with the ultimate goal of minimizing the most loaded link in the network. The proposed system is compatible with any network running a link-state intradomain routing protocol (e.g., OSPF, IS-IS, etc)\footnote{The source code and all data needed to reproduce our experiments is available at \url{https://github.com/BNN-UPC/Papers/wiki/MARL-GNN-TE}}.

Unlike previous ML-based proposals for TE (e.g., \cite{valadarsky2017learning,xu2018experience,ding2020packet}), the combination of MARL and GNN allows us to handle topologies of various sizes and structures in a distributed fashion; and more importantly, to achieve combinatorial generalization over the information exchanged by agents in the network, which is naturally represented as a \mbox{graph~\cite{battaglia2018relational,suarez2019challenging}}. 

At the time of this writing, the most advanced proposals for TE are based on refined optimization algorithms, such as Constraint Programming~\cite{hartert2015declarative}, Local Search~\cite{gay2017expect}, Mixed Integer Linear Programming~\cite{bhatia2015optimized} or Column Generation~\cite{jadin2019cg4sr}. All these solutions offer significant performance improvements with respect to traditional routing strategies, such as shortest path routing or load balancing. In our evaluation, we focus on DEFO~\cite{hartert2015declarative} as a benchmark for comparison, which is arguably among the best performing and most advanced TE solutions available~\cite{gay2017repetita}. DEFO is a sophisticated proposal that represents the result of decades of research in TE. One of the main challenges in TE optimization is to reduce the dimensionality of the vast search space. To this end, DEFO proposes iddlepoint Routing (MR), a smart abstraction of routing inspired by Segment Routing~\cite{filsfils2015segment}. With this approach, and leveraging the SDN paradigm to implement a fully centralized optimization algorithm, the authors are able to optimize a network with close-to-optimal performance within few minutes, which allows for near real-time operation.

In our experimental evaluation, we benchmark the proposed MARL system against DEFO across a wide variety of network scenarios. We first evaluate the performance of our solution in a set of real-world network topologies with realistic traffic matrices, and report comparable performance to DEFO. Then, we analyze the execution time (speed) of both solutions and show that our MARL system only requires few seconds to optimize a network, while DEFO operates at the scale of minutes~\cite{hartert2015declarative}. This improvement in terms of execution time is a result of the inherent distributed nature of our MARL system, which allows us to share the computation among the network devices without requiring any centralized entity. This enables for sub-minute real-time operation with comparable performance to the state-of-the-art in TE.

\section{Network Traffic Engineering Scenario} \label{sec:scenario}

In this section, we describe the network scenario considered in this paper. Particularly, we address the intradomain TE problem, where network traffic is measured and routed to optimize the network resources. Typically, IP networks run link-state Interior Gateway Protocols (IGP), such as Open Shortest Path First (OSPF)~\cite{RFC2328}, that choose paths using the Dijkstra's algorithm over some \mbox{pre-defined} link weights. 

TE can be achieved through different approaches. Network operators use commercial tools~\cite{cisco2013mate,juniper} to fine-tune link weights. Other mechanisms propose to add extra routing entries~\cite{sridharan2005achieving} or end-to-end tunnels (e.g., RSVP-TE~\cite{minei2005mpls}) to perform source-destination routing, expanding the solution space. In the literature, we can find a wide range of proposed architectures and algorithms for TE~\cite{huang2018survey}.

Our proposed ML solution is a fully distributed architecture that optimizes link weights (similar to existing commercial solutions~\cite{cisco2013mate,juniper}) and interfaces with standard OSPF. It does not require any changes to OSPF, while it can be implemented with a software update on the routers where it is deployed. Relying on well-known link-state routing protocols, such as OSPF, also offers the advantage that the network is easier to manage compared to finer-grained alternatives, such as flow-based routing~\cite{xu2011link}. 

In what follows, we describe the three operational steps of our solution (see Fig.~\ref{fig:implementation}):

\textbf{1) Traffic Measurement:} In step \circled{1}\hspace{-3pt}, a traffic measurement platform deployed over the network identifies a new Traffic Matrix (TM). This new TM needs to be communicated to all participating routers, which upon reception will start the next step and optimize the routing for this TM. We leave out of the scope of this paper the details of this process, as TM estimation is an extensive research field with many established proposals. For instance, this process can be done periodically (e.g, each 5-10 minutes as in \cite{fortz2000internet}), where the TM is first estimated and then optimized. Some proposals trigger the optimization process when a relevant change is detected in the TM \cite{benson2011microte}, while others use prediction techniques to optimize it in advance \cite{luo2013dsox}. Finally, some real-world operators make estimates considering their customers' subscriptions and operate based on a static TM. Our proposal is flexible and can operate with any of these approaches.

\textbf{2) Proposed MARL+GNN TE optimization system:} Upon reception of the TM, routers run the MARL+GNN optimization process (step~\circled{2}\hspace{-2pt}), which eventually computes the per-link weights that optimize OSPF routing in the subsequent step. Particularly, we set the goal to minimize the maximum link load (\mbox{\emph{MinMaxLoad}}), which is a classic TE goal in carrier-grade networks~\cite{azar2004optimal, fortz2004increasing,hartert2015declarative}. This problem is known to be NP-hard, and even good settings of the weights can deviate significantly from the optimal configuration~\cite{xu2011link,fortz2004increasing}. Our MARL optimization system is built using a distributed Graph Neural Network (GNN) that exchanges messages over the physical network topology. Messages are sent between routers and their directly attached neighbors. The content of such messages are \emph{hidden states} that are produced and consumed by artificial neural networks and do not have a human-understandable \emph{meaning}. The GNN makes several message iterations and, during this phase, local configuration of the router remains unchanged, thus having no impact on the current traffic. More details about the inner workings, performance, communication overhead, and computational cost can be found in Sections~\ref{sec:architecture} and~\ref{sec:results}.

\textbf{3) OSPF convergence:} Finally, step~\circled{3}\hspace{-4pt} is the standard OSPF convergence process based on the new per-link weights computed by the MARL+GNN system. Specifically, each agent has computed the optimal weigths for its locally attached links. For OSPF to recompute the new forwarding tables, it needs to broadcast the new link weights; this is done using the standard OSPF Link-State Advertisements (LSAs) \cite{RFC2328}. Once the routers have an identical view of the network, they compute locally their new forwarding tables, and traffic is routed following the optimization goal. Convergence time of OSPF is a well-studied subject. For instance, routing tables can converge in the order of a few seconds in networks with thousands of links \cite{basu2001stability}.

\section{Background} \label{sec:back}

The solution proposed in this paper incorporates two ML-based mechanisms: GNNs and MARL. In this section, we provide some background on these technologies.

\subsection{Graph Neural Networks} \label{subsec:back-GNN}

GNNs are a novel family of neural networks designed to operate over graphs. They were introduced in \cite{scarselli2008graph}, and numerous variants have been developed since \cite{wu2020comprehensive}. In their basic form, they consist in associating some initial states to the different elements of an input graph, and combine them considering how these elements are connected in that graph. The resulting state representations, which now may encode some topological awareness, are then used to produce the final output of the GNN, which can be at the level of graph elements, or at a global graph level. 

In particular, we will focus on Message Passing Neural Networks (MPNN)~\cite{gilmer2017neural}, which is a well-known type of GNN whose operation is based on an iterative message-passing algorithm that propagates information between the selected elements of the graph --for simplicity, let us assume that we consider as elements the nodes of such graph. First, each node $v$ initializes its hidden state $h_v^0$ using some initial features already included in the input graph. At every message-passing step $k$, each node $v$ receives via messages the current hidden state of all the nodes in its neighborhood $\mathcal{B}(v)$, and processes them individually by applying a message function \textit{m(·)} together with its own internal state $h_v^k$. Then, the processed messages are combined by an aggregation function \textit{a(·)}:
\begin{equation}
M_v^k = a( \{ m(h_{v}^k, h_{i}^{k}) \}_{i \in \mathcal{B}(v)})
\label{eq:message_function}
\end{equation}
Finally, an update function \textit{u(·)} is applied to each node $v$; taking as input the aggregated messages $M_{v}^{k}$ and its current hidden state $h_v^k$, it outputs a new hidden state for the next step ($k+1$):
\begin{equation}
h_{v}^{k+1} = u(h_{v}^{k}, M_{v}^{k}).
\vspace{0.15cm}
\label{eq:update_function}
\end{equation}
After a certain number of message passing steps $K$, a readout function \textit{r(·)} takes as input the final node states $h_v^K$ to produce the final output of the GNN model. This readout function can predict either features of individual elements (e.g., a node's class) or global properties of the graph.

We note that a MPNN model generates \textit{a single set of message, aggregation, update, and readout functions that are replicated at each selected graph element}. This means that these functions should be generic and flexible enough to adapt their behaviour to different scenarios, which is why they are usually modeled as traditional neural networks --specially fully connected and recursive neural networks, with the only exception of the aggregation function that is commonly an element-wise summation.

\subsection{(Multi-Agent) Reinforcement Learning} \label{subsec:back-RL}

In the standard Reinforcement Learning (RL) setting\cite{bertsekas1996neuro}, an agent interacts with the environment in the following way: at each step $t$, the agent selects an action $a_t$ based on its current state $s_t$, to which the environment responds with a reward $r_t$ and then moves to the next state $s_{t+1}$. This interaction is modeled as an episodic, time-homogeneous Markov Decision Process (MDP) $(\Sspace, \A, r, P, \gamma)$, where $\Sspace$ and $\A$ are the state and action spaces, respectively; $P$ is the transition kernel, $s_{t+1} \sim P(\cdot | s_t,a_t)$; $r_t$ represents the immediate reward given by the environment after taking action $a_t$ being in state $s_t$; and $\gamma \in (0,1]$ is the discount factor used to compute the return $G_t$, defined as the --discounted-- cumulative reward from a certain time-step $t$ to the end of the episode $T$: $G_t = \sum_{t=0}^T \gamma^t r_t$. The behavior of the agent is described by a policy $\pi: \mathcal{S} \to \mathcal{A}$, which maps each state to a probability distribution over the action space, and the goal of an RL agent is to find the optimal policy in the sense that, given any considered state $s\in \Sspace$, it always selects an action that maximizes the expected return~$\hat{G}_t$.

In this work, we focus on model-free Policy Gradient Optimization methods~\cite{sutton2018reinforcement}, where the agent learns an explicit policy representation $\pi_\theta$ with some parameters~$\theta$ --typically a neural network. In most cases, during the training process, they involve learning as well a function approximator $V_\phi(s)$ of the state value function $V^{\pi_\theta}(s)$, defined as the expected discounted return from a given state $s$ by following policy $\pi_\theta$: 
\begin{equation}
    V^{\pi_\theta} (s) = \E _{\pi_\theta} \left[ G_t | s_t=s \right]
\end{equation}
This defines the so-called Actor-Critic family of Policy Gradient algorithms~\cite{sutton2018reinforcement}, where actions are selected from the function that estimates the policy (i.e., the actor), and the training of such policy is guided by the estimated value function to assess the consequences of the actions taken (i.e., the critic). Our solution is precisely based on an Actor-Critic method named Proximal Policy Optimization (PPO)~\cite{schulman2017proximal}, which offers a favorable balance between reliability, sample complexity, and simplicity; we refer the reader to the original paper~\cite{schulman2017proximal} for further details.

Contrary to a single-agent RL setting, in a Multi-Agent Reinforcement Learning (MARL) framework there is a set of agents $\mathcal{V}$ interacting with a common environment that have to learn how to cooperate to pursue a common goal. Such a setting is generally formulated as a Decentralized Partially Observable MDP (Dec-POMDP)~\cite{foerster2018deep} where, besides the global state space $\Sspace$ and action space $\A$, it distinguishes local state and action spaces for every agent --i.e., $\Sspace_v$ and $\A_v$ for $v \in \mathcal{V}$. At each time step $t$ of an episode, each agent may choose an action $a^v_t \in \A_v$ based on local observations of the environment encoded in its current state $s^v_t \in \Sspace_v$. Then, the environment produces individual rewards $r_t^v$ (and/or a global one $r_t$), and it evolves to a next global state $s_{t+1}\in \Sspace$ --i.e., each agent $v$ turn into the following state $s^v_{t+1} \in \Sspace_v$. Typically, a MARL system seeks for the optimal global policy by learning a set of local policies $\{\pi_{\theta_v}\}_{v \in \mathcal{V}}$. For doing so, most state-of-the-art MARL solutions implement traditional (single-agent) RL algorithms on each distributed agent, while incorporating some kind of cooperation mechanism between them~\cite{foerster2018deep}. The standard approach for obtaining a robust decentralized execution, however, is based on a centralized training where extra information can be used to guide agents' learning \cite{oliehoek2008}.

\section{MARL+GNN Architecture} \label{sec:architecture}

The TE scenario described in Section \ref{sec:scenario} is implemented through a MARL architecture that, thanks to the use of GNN, yields good generalization properties over networks~\cite{battaglia2018relational,suarez2019challenging}. We will first introduce our generic MARL+GNN framework, which is especially designed for distributed networking tasks, and then we will provide details on how it is adapted to the intradomain TE use case addressed in this paper.

\subsection{Framework Formulation} \label{subsec:framework}

We model a networked MARL environment as a graph \mbox{$\mathcal{G} = (\Nagents, \mathcal{E})$}, with some nodes $n \in \Nagents$ and edges $e \in \mathcal{E}$, where a set of agents $\mathcal{V}$ control some of the graph entities (nodes or edges). Our architecture extends the single-agent PPO method~\cite{schulman2017proximal} to accommodate the distributed multi-agent environment. 

In contrast to the standard MARL setting described in Section~\ref{subsec:back-RL}, where a policy $\pi_{\theta_v}$ is learned for each agent $v \in \mathcal{V}$, we propose to directly learn a global policy $\pi_\theta: \Sspace \rightarrow \A$ in a distributed fashion over the global state and action spaces, defined as the joint and union of the respective agents' local spaces -- i.e., $\Sspace = \prod_{v\in \mathcal{V}} \Sspace_v$ and $\A = \bigcup_{v\in \mathcal{V}} \A_v$. This allows us to formulate the problem as a classic MDP, thus avoiding the more complex Dec-POMDP scenario.

An important novelty of our design is that all agents $v\in\mathcal{V}$ are able to internally construct the global policy representation $\pi_\theta$ mainly through message communications with their direct neighboring agents $\mathcal{B}(v)$ and their local computations. Thus, we no longer need a centralized entity responsible for collecting and processing all the global information together to provide~$\pi_\theta$. Such a decentralized, message-based generation of the global policy can be achieved by modeling the actor with a MPNN (see Sec.~\ref{subsec:back-GNN}), so that $\pi_\theta$ is now encoded as a GNN rather than as a classical, non-relational neural network (e.g., fully-connected NN). In particular, this implies that all agents $\mathcal{V}$ deployed in the network are actually elements of a larger-scale mechanism --orchestrated by the MPNN-- that requires them to perform regular message exchanges with their neighbors. Algorithm \ref{alg:pipeline} summarizes the full execution pipeline of our solution.

\setlength{\algomargin}{1.5em}
\SetAlCapHSkip{0em}
\begin{algorithm}[!t]
\caption{MARL+GNN execution pipeline.} \label{alg:pipeline}
\DontPrintSemicolon
\SetKwComment{CustomComment}{\#}{}
\SetKwInput{KwIn}{\hspace{-1.5em} Input}
\SetKwInput{KwOut}{\hspace{-1.5em} Output}
\SetKwInOut{Require}{\hspace{-1.5em} Require}
\SetKw{KwBy}{by}
\Require{A graph $\mathcal{G} = (\Nagents, \mathcal{E})$ with a set of agents $\mathcal{V}$, MPNN trained parameters $\theta=\{\theta_i\}_{i\in \{m,a,u,r\}}$}
\KwIn{Initial graph configuration $X^0_{\mathcal{G}}$, episode length $T$, number of message passing steps $K$}
    \textnormal{Agents initialize their states $s_v^0$ based on $X^0_{\mathcal{G}}$}\;
    \For{$t \leftarrow 0$ \KwTo $T$}{
        \textnormal{Agents initialize their hidden states }$h_v^0 \leftarrow (s_v^t, 0, \dots , 0)$\;
        \For{$k \leftarrow 0$ \KwTo $K$}{
            \textnormal{Agents share their current hidden state $h_v^k$ to neighboring agents $\mathcal{B}(v)$}\;
            \textnormal{Agents process the received messages $M_v^{k} \leftarrow a_{\theta_a}( \{ m_{\theta_m}(h_{v}^k, h_{\mu}^k) \}_{\mu \in \mathcal{B}(v)})$}\;
            \textnormal{Agents update their hidden state $h_{v}^{k+1}\leftarrow u(h_{v}^{k}, M_v^{k})$}\;
        }
        \textnormal{Agents compute their actions' logits $\{ \textnormal{logit}_v (a) \}_{a \in \A_v} \leftarrow r_{\theta_r}(h_v^K)$}\;
        \textnormal{Agents receive the actions' logits of the rest of agents and compute the global policy $\pi_\theta \leftarrow \textnormal{CategoricalDist} \left( \{ \{ \textnormal{logit}_v (a) \}_{a \in \A_v} \}_{v \in \mathcal{V}} \right)$}\;
        \textnormal{Using the same probabilistic seed, agents sample an action $a_t \in \A_{v'}$, for $v' \in \mathcal{V}$, from policy $\pi_\theta$}\;
        \textnormal{Agent $v'$ executes action $a_t$, and the environment updates the graph configuration $X^{t+1}_{\mathcal{G}}$}\;
        \textnormal{Agents update their states $s_v^{t+1}$ based on $X^{t+1}_{\mathcal{G}}$}\;
    }
\KwOut{New graph configuration $X^{*}_{\mathcal{G}}$ that optimizes some pre-defined objective or metric}
\end{algorithm}

Inherently, at each step $t$ of the episodic MDP, the MPNN-driven process of estimating the policy $\pi_\theta(\cdot|s_t)$ first requires engineering a meaningful hidden state $h_v$ for each agent $v\in\mathcal{V}$. Each hidden state $h_v$ basically depends on the hidden representations of the neighboring agents $\mathcal{B}(v)$, and its initialization $h_v^0$ is a function of the current agent state $s^t_v$, which is in turn based on some pre-defined internal agent features $x_v^t$. Those representations are shaped during $K$ message-passing steps, where hidden states are iteratively propagated through the graph via messages between direct neighbors. In particular, successive hidden states $h_v^k$, where $k$ refers to the message-passing step, are computed by the message, aggregation and update functions of the MPNN, as described in Section \ref{subsec:back-GNN}. 

Once agents generate their final hidden representation, a readout function --following the MPNN nomenclature-- is applied to each agent to finally obtain the global policy distribution~$\pi_\theta$. Particularly, in our system the readout is divided into two steps: first, each agent $v\in\mathcal{V}$ implements a local readout that takes as input the final representation $h^K_v$, and produces as output the unnormalized log probability (i.e., logit) of every possible action in the agent's space $\A_v$. The second and last step involves a communication layer that propagates the logits among agents, so that all of them can internally construct the global policy $\pi_{\theta}$ for the overall network state $s_t = \prod_{v\in\mathcal{V}}s^t_v$. To ensure that all the distributed agents sample the same actions along the message-passing process $a^t_{v'} \sim \pi_{\theta}(\cdot | s_t)$, $v' \in \mathcal{V}$, they share a common seed before initiating this process. Consequently, only the agent $v'$ whose action has been selected does execute an action at each time-step $t$.

For each step of an episode -- of length $T$ -- our solution runs the MPNN-based actor model described above, after applying the action selected in the previous step. Note that each action could modify one or several agent's internal states, which would vary their hidden state initializations, hence leading to a completely new optimization process. During training, each agent stores the global trajectory $\{s_t, a_t, s_{t+1}\}_{t=0}^T$, from which they can learn the configuration that leads to better global performance at the end of the episode.

One especial characteristic of the proposed system with respect to common MARL settings emerges from the internal implementation of a MPNN model. As a result, rather than having independent functions for each agent as in the standard MARL setting~\cite{foerster2018deep}, in our system all agents implement the same functions (i.e., message, aggregation, update, and readout) with the same parameters $\theta$. Hence, all agents run exactly the same processing pipeline, and their outcome depends on both their initialization and the local information received from their neighbors.

Indeed, our solution produces a single \emph{universal agent} implementation that builds upon the inner functions of the MPNN, which are jointly learned during training across all the agents instances in the network (see Sec.~\ref{subsec:back-GNN} for more details). Thus, after training, each agent $v \in \mathcal{V}$ can be interpreted as a replica of this \emph{universal agent} that behaves based on its local environment. This so-called \emph{parameter sharing} feature provides compelling generalization and scalability properties, which can be beneficial to effectively deploy the solution in networks with topologies of different size and structure, not necessarily seen during the training phase~\cite{battaglia2016interaction, suarez2019challenging}.

\subsection{Application to Traffic Engineering} \label{subsec:routing}

A straightforward approach to map the previously described MARL+GNN setting to the intradomain TE problem is to associate agents to each element of the physical topology --i.e., given a network topology, each agent can control individually the configuration of a network device, or some configuration parameters of a link connecting two devices. In the network scenario described in Section \ref{sec:scenario}, we consider that each agent controls a link (i.e., $\mathcal{V} = \mathcal{E}$). In practice, these link-based agents are executed in the adjacent device of the link (e.g., router). Figure \ref{fig:message_passing} shows a visual representation of our distributed MARL+GNN system adapted to the TE use case, particularly with the goal of minimizing the most loaded link~\cite{azar2004optimal, fortz2004increasing,hartert2015declarative}. Taking this figure as a reference, we describe the particular adaptations of our MARL framework to be applied to the selected intradomain TE scenario:

\begin{figure}[!t]
\centering
    \includegraphics[width=\columnwidth]{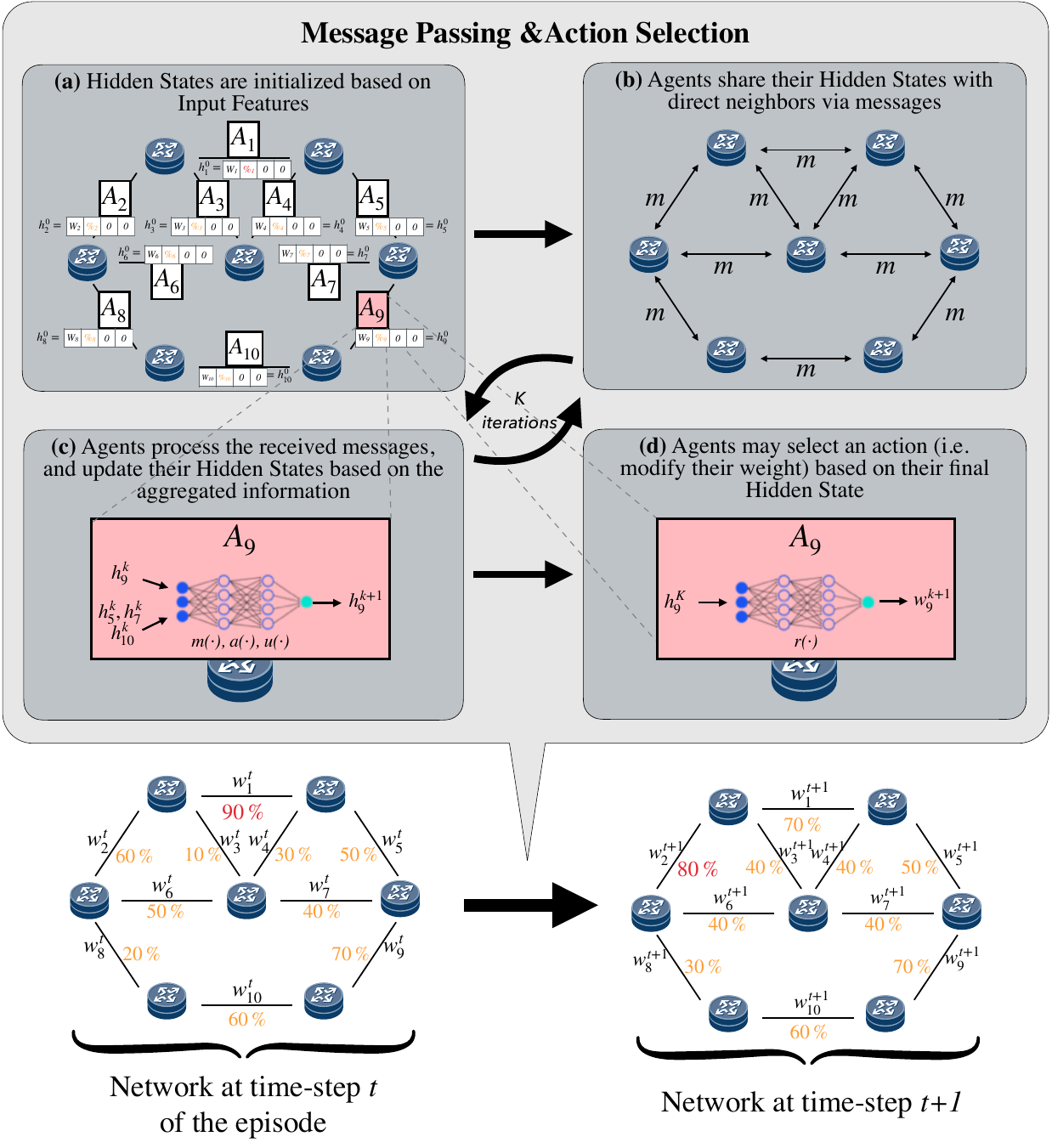}
  \caption{Description of the message passing and action selection process of our MARL+GNN solution in a time-step. The full procedure is repeated $T$ times, which is the pre-defined episode length.}
  \label{fig:message_passing}
  \vspace*{-0.2cm}
\end{figure} 

\subsubsection{Environment}

In the traditional MDP setting, we consider episodes of a fixed number of time-steps $T$. 
At the beginning of each episode, the environment provides with a set of traffic demands between all source-destination pairs (i.e., an estimated traffic matrix~\cite{fortz2000internet, benson2011microte, luo2013dsox}). Each link $e \in \mathcal{E}$ has an associated capacity $c_e$, and it is initialized with a certain link weight $w_e^0$. These link weights are in turn used to compute the routers' forwarding tables (via standard OSPF convergence). Each agent $v_e \in \mathcal{V}$ has access to its associated link features, which in our case are the current weight, its capacity, and also the estimated traffic matrix and the weights of the other links. This can be achieved with standard procedures in OSPF environments (see Sec.~\ref{sec:scenario}).

\subsubsection{State Space and Message Passing}

At each time-step~$t$ of an episode, each link-based agent $v_e \in \mathcal{V},$ feeds its MPNN module with its input features $x_e^t$ to generate its respective initial hidden state $h_e^0$ (Figure~\ref{fig:message_passing}-a). In particular, agents consider as input features the current weight $w_e^t$ and the utilization $u_e^t$ $[0,1]$ of the link, and construct their initial link hidden representations $h_e^0$ as a fixed-size vector where the first two components are the input features and the rest is zero-padded. Note that the link utilization can be easily computed by the agent with the information of the estimated traffic matrix and the global link weights locally maintained. Then, the algorithm performs \textit{K} message-passing steps \mbox{(Figures~\ref{fig:message_passing}-b and 2-c)}. At each step $k$, the algorithm is executed in a distributed fashion over all the links of the network. Particularly, for each link, the corresponding agent receives the hidden states of its neighboring agents (i.e., adjacent links), and combines them individually with its own state $h_e^k$ ($message$ function), using a fully-connected NN. Then, all the messages computed in each link (with its neighbors) are aggregated using an element-wise sum, producing an aggregated message $M_e^k$. Afterwards, another fully-connected NN is used as $update$ function, which combines the link hidden state $h_e^k$ with the new aggregated information $M_e^k$, and produces a new hidden state representation for that link ($h_e^{k+1}$). As mentioned above, this process is repeated \textit{K} times, leading to some final link hidden state representations $h_e^K$. In short, during this K message-passing process, agents increasingly transform their initial link hidden states (initialized with the link weight and utilization) based on their local communications with adjacent agents (i.e., links).

\subsubsection{Action Space}

In our TE approach, the possible action of each agent $e\in \mathcal{E}$ is to modify the weight of its associated link $w_e$. Due to \emph{parameter sharing}, all of them share the same action space $\A_e$. In our specific implementation each agent has only one possible action at each time-step: to increase the link weight in one unit. Note that the same agent can increment more than once its weight along an episode, thus providing enough expressiveness to generate potentially any combination of link weights at the end of the episode. In particular, the agent's action selection (Figure \ref{fig:message_passing}-d) is done as follows: first, every agent applies a local readout function --implemented with a fully-connected NN-- to its final hidden state $h_e^K$, from which it obtains the global logit estimate of choosing its action (i.e., increase its link weight) over the actions of the other agents. Then, as previously described in Section \ref{subsec:framework}, these logits are shared among agents in the network, so that each of them can construct the global policy distribution $\pi_\theta$. By sharing the same probabilistic seed (from the beginning of the episode), all the agents sample locally the same action $a_{e'}^t$ from $\pi_\theta$, thus selecting the agent $v_{e'}\in \mathcal{V}$ that will increase its weight and, consequently, each agent increases by one the weight of the selected link in its internal global state copy, which is then used to initialize its hidden state representation in the next time-step $t+1$; particularly, to compute the new link utilization $u_e^{t+1}$ under this new weight setting.

\subsubsection{Reward Function}

During training, a reward function is computed at each step $t$ of the optimization episode. In our case, as the optimization goal is to minimize link congestion, we define the reward $r_t$ as the difference of the global maximum link utilization between steps $t$ and $t+1$. Note that this reward can be computed locally in each agent from its global state copy, which is incrementally updated with the new actions applied at each time-step $a_{e'}^t$.

\subsection{Training Phase}

Formally, during training the goal is to optimize the parameters $\{\theta, \phi\}$ so that:
\begin{itemize}
    \item The previously described GNN-based actor $\pi_\theta$ becomes a good estimator of the optimal global policy;
    \item The critic $V_\phi$ learns to approximate the state value function of any global state\footnote{The critic is exclusively used for training, it is no longer needed at runtime. 
    We have implemented it as an independent link-based MPNN, similar to the actor, in order to exploit the relational reasoning provided with GNNs. However, other alternative designs would be valid as well.
    }.
\end{itemize}

In particular, the training pipeline is done as follows: An episode of length $T$ is generated by following the current policy $\pi_\theta$, while at the same time the critic's value function $V_\phi$ evaluates each visited global state; thus, the episode defines a trajectory $\{s_t,a_t,r_t,p_t,V_t,s_{t+1}\}_{t=0}^{T-1}$, where $p_t = \pi_\theta(a_t | s_t)$ and $V_t := V_\phi (s_t)$. When the episode ends, this trajectory is used to update the model parameters --through several epochs of minibatch Stochastic Gradient Descent-- by maximizing the global PPO objective $L^{PPO}(\theta, \phi)$ described in \cite{schulman2017proximal}.

\section{Evaluation} \label{sec:results}

In this section we make an extensive set of experiments --over real-world network topologies-- to evaluate the proposed MARL+GNN architecture (Sec.~\ref{sec:architecture}). We particularly focus on comparing the proposed solution with DEFO~\cite{hartert2015declarative}, which is arguably among the best performing and most advanced TE solutions available at the time of this writing~\cite{gay2017repetita}.

\subsection{Experimental Setup}

Along the evaluation section, we consider three real-world network topologies for training and evaluation of our model: 42-link NSFNet, 54-link GBN, and 72-link GEANT2~\cite{rusek2019unveiling}. The length $T$ of the training and evaluation episodes is pre-defined, and it varies from 100 to 200 steps, depending on the network topology size (see more details later in Sec.~\ref{subsec:steps}). At the beginning of each episode, the link weights are randomly selected as an integer in the range $[1,4]$, so our system is evaluated over a wide variety of scenarios with random routing initializations. From that point on, at each step of an episode a single agent can modify its weight by increasing it in one unit, thus chaining the selected actions on the T time-steps of an episode.

Taking~\cite{andrychowicz2021what} as a reference for defining the hyperparameters' values of the solution, we ran several grid searches to appropriately fine-tune the model. The implemented optimizer is Adam with a learning rate of $3 \cdot 10^{-4}$, $\beta$=$0.9$, and $\epsilon$=$0.9$. Regarding the PPO setting, the number of epochs for each training episode is set to $3$ with batches of size $25$, the discount factor $\gamma$ is set to $0.97$, and the clipping parameter to $0.25$. We implement the Generalized Advantage Estimate (GAE), to estimate the advantage function with $\lambda$=$0.9$. In addition, we multiply the critic loss by a factor of $0.5$, and we implement an entropy loss weighted by a factor of $0.001$. Finally, links' hidden states $h_e$ are encoded as 16-element vectors, and in each MPNN forward propagation $K$=$8$ message passing steps are executed.

We consider two different traffic profiles: \textit{(i)} uniform distribution of source-destination traffic demands, and \textit{(ii)} traffic distributions following a gravity model~\cite{roughan2005simplifying}, which produces more realistic Internet traffic matrices. For each set of experiments, the training process of our MARL+GNN system took about 24 hours running in a machine with a single CPU of 2.20 GHz ($\sim$1M training steps).

\subsection{Baselines} \label{subsec:baselines}

This section describes the baselines we use to benchmark our MARL+GNN system in our experiments. We particularly consider two well-known TE alternatives:
\begin{itemize}
	\item Default OSPF: We consider the routing configuration obtained by applying the OSPF protocol with the common assumption that link weights are inversely proportional to their capacities. We consider traffic splitting over multiple paths (OSPF with ECMP), which is a standard recommended best practice.
	\item Declarative and Expressive Forwarding Optimizer (DEFO)~\cite{hartert2015declarative}: A centralized network optimizer that translates high-level goals of operators into network configurations in real-time (in the order of minutes). DEFO starts from a routing configuration already optimized with a commercial TE tool \cite{cisco2013mate}, and it uses Constraint Programming~\cite{rossi2006handbook} and Segment Routing~\cite{filsfils2015segment} to further optimize it. To this end, DEFO reroutes traffic paths through a sequence of middlepoints, spreading their traffic over multiple ECMP paths. DEFO obtains close-to-optimal performance considering several network optimization goals, one of them being our intradomain TE goal of minimizing the most loaded link. We use the code publicly shared by the authors of DEFO\footnote{\url{https://sites.uclouvain.be/defo/}}. For the sake of comparison, we also use OSPF-ECMP in the evaluation of our system (MARL+GNN), although it can also operate in scenarios without ECMP support.
\end{itemize}

\subsection{Performance Evaluation over different Traffic Matrices} \label{subsec:res-TM}

In this subsection we present the results of our first experiment, which evaluates the performance of our proposed MARL solution over traffic matrices that have not been seen during the training process. More in detail, we consider a fixed network topology and a set of traffic matrices; then our model is trained in that single topology using a subset of traffic matrices, and finally the trained system is evaluated over a different set with unseen traffic. 

In particular, we analyze two different traffic profiles (uniform and gravity model), each of them in two network topologies (NSFNet and GEANT2). In total we run four independent experiments, one for each combination of traffic profile and topology. At each experiment, we stop the training when the system has observed around $100$ different traffic matrices (TM), and the model is evaluated over $100$ new TMs. During training, TMs change every $50$ training episodes.

\begin{figure}[!t]
    \begin{subfigure}[]{0.495\columnwidth}
	\includegraphics[width=1.0\linewidth,height=3.5cm]{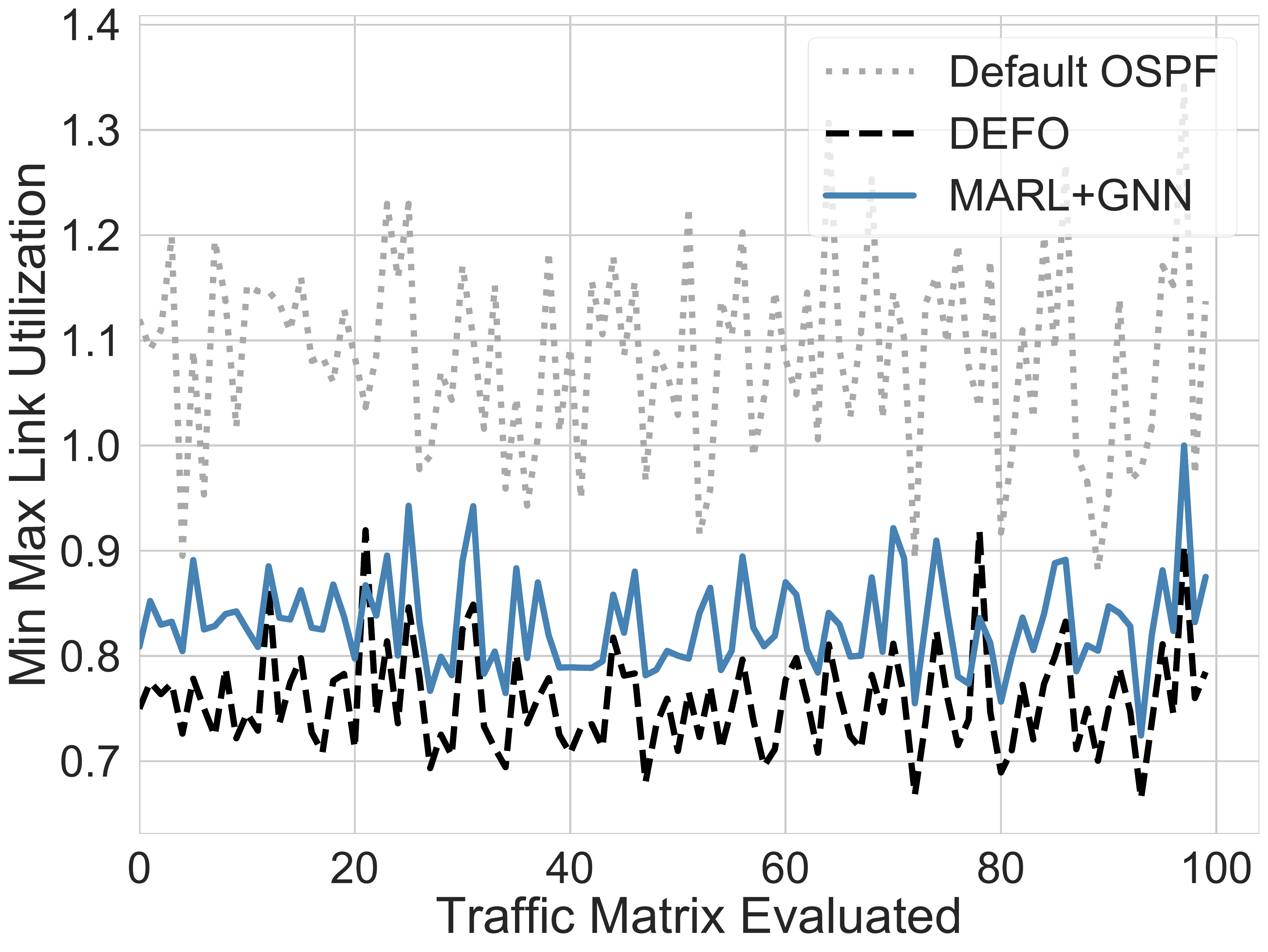}
        \caption{\textit{MinMaxLoad} NSFNet}
	\label{subfig:raw_NSFNet_uniform}
    \end{subfigure}
    \begin{subfigure}[]{0.495\columnwidth}
	\includegraphics[width=1.0\linewidth,height=3.5cm]{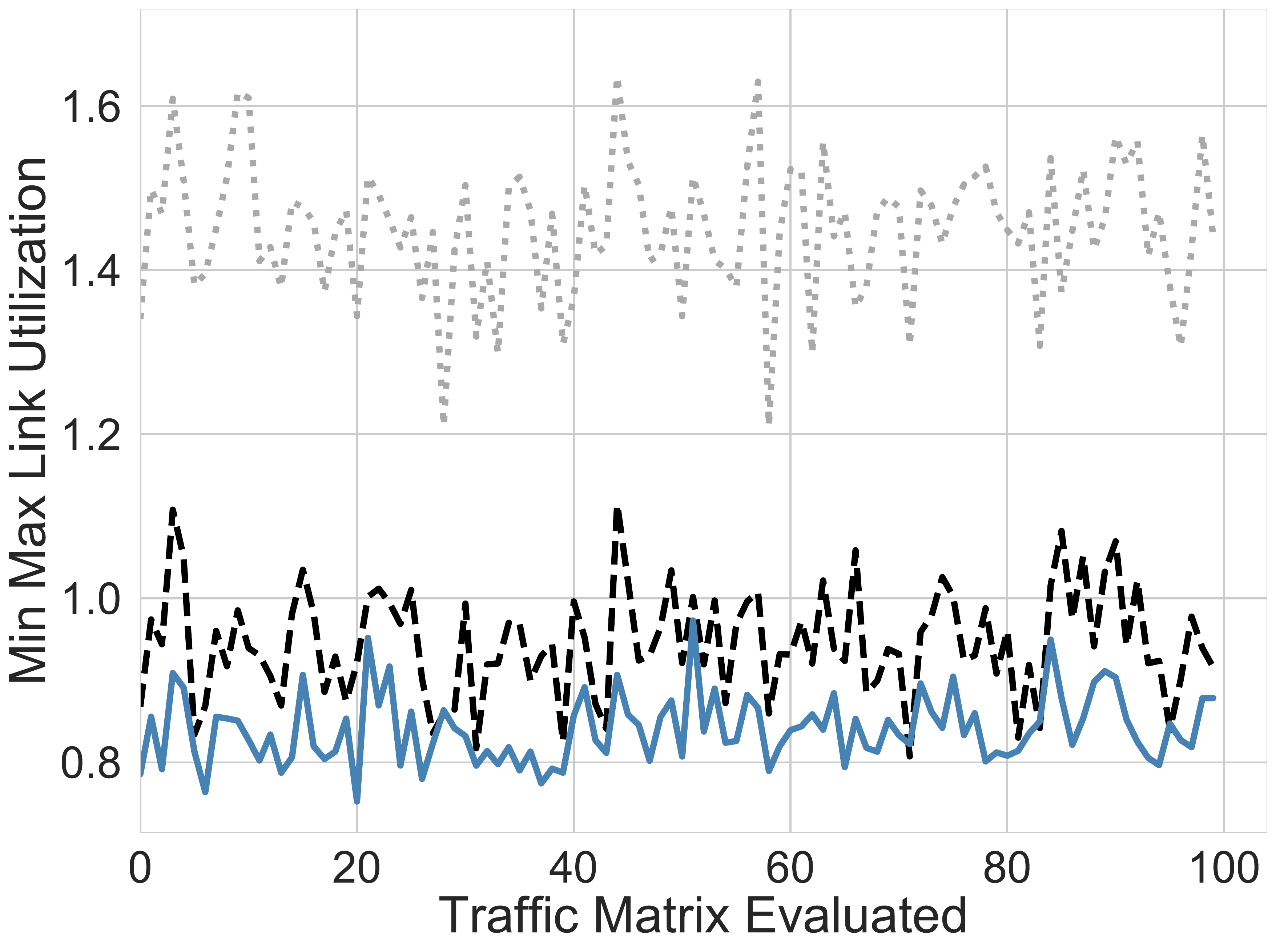}
        \caption{\textit{MinMaxLoad} GEANT2}
	\label{subfig:raw_GEANT2_uniform}
    \end{subfigure}
    \begin{subfigure}[]{0.495\columnwidth}
	\includegraphics[width=1.0\linewidth,height=3.5cm]{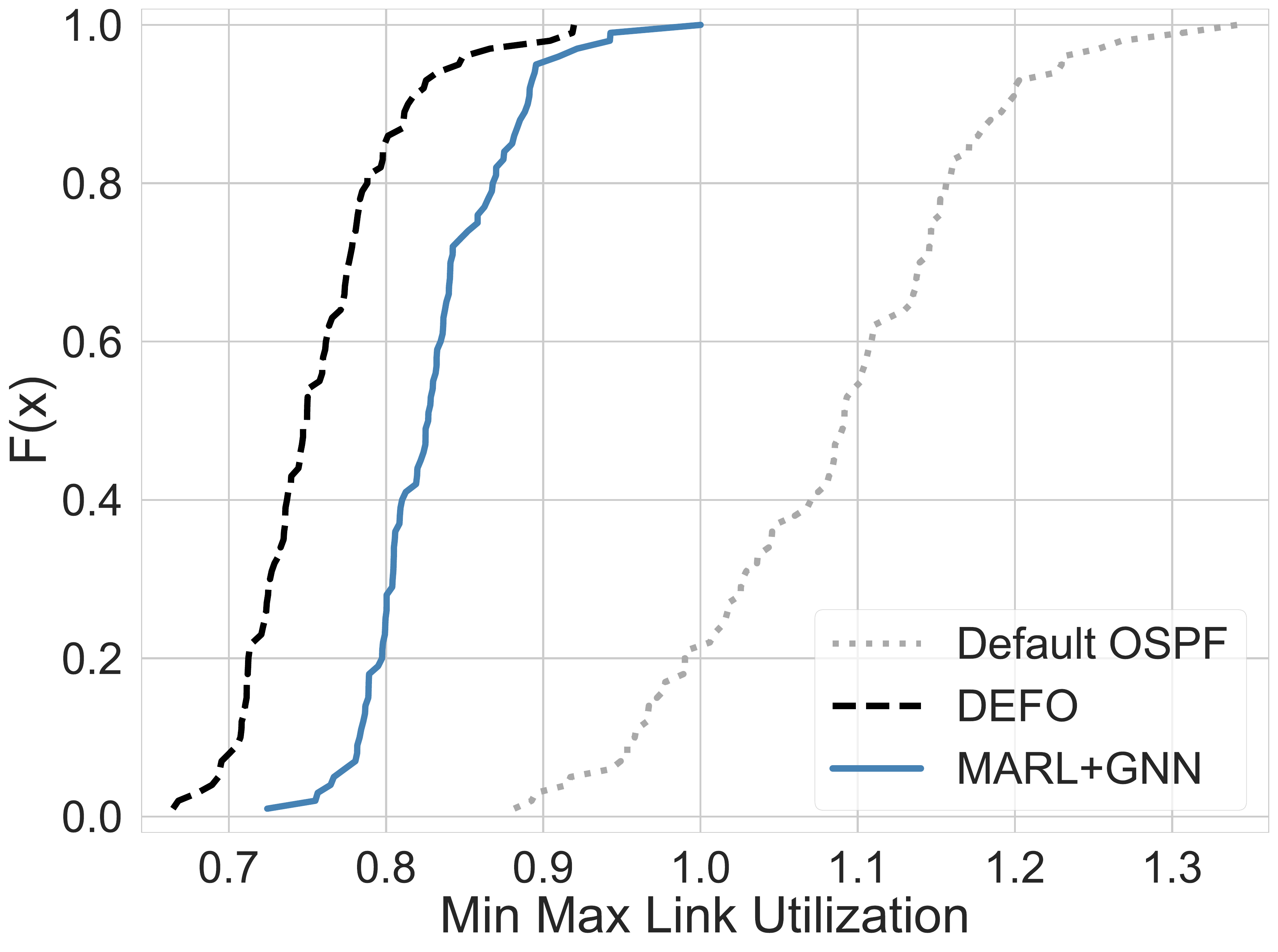}
        \caption{CDF NSFNet}
	\label{subfig:CDF_NSFNet_uniform}
    \end{subfigure}
    \begin{subfigure}[]{0.495\columnwidth}
	\includegraphics[width=1.0\linewidth,height=3.5cm]{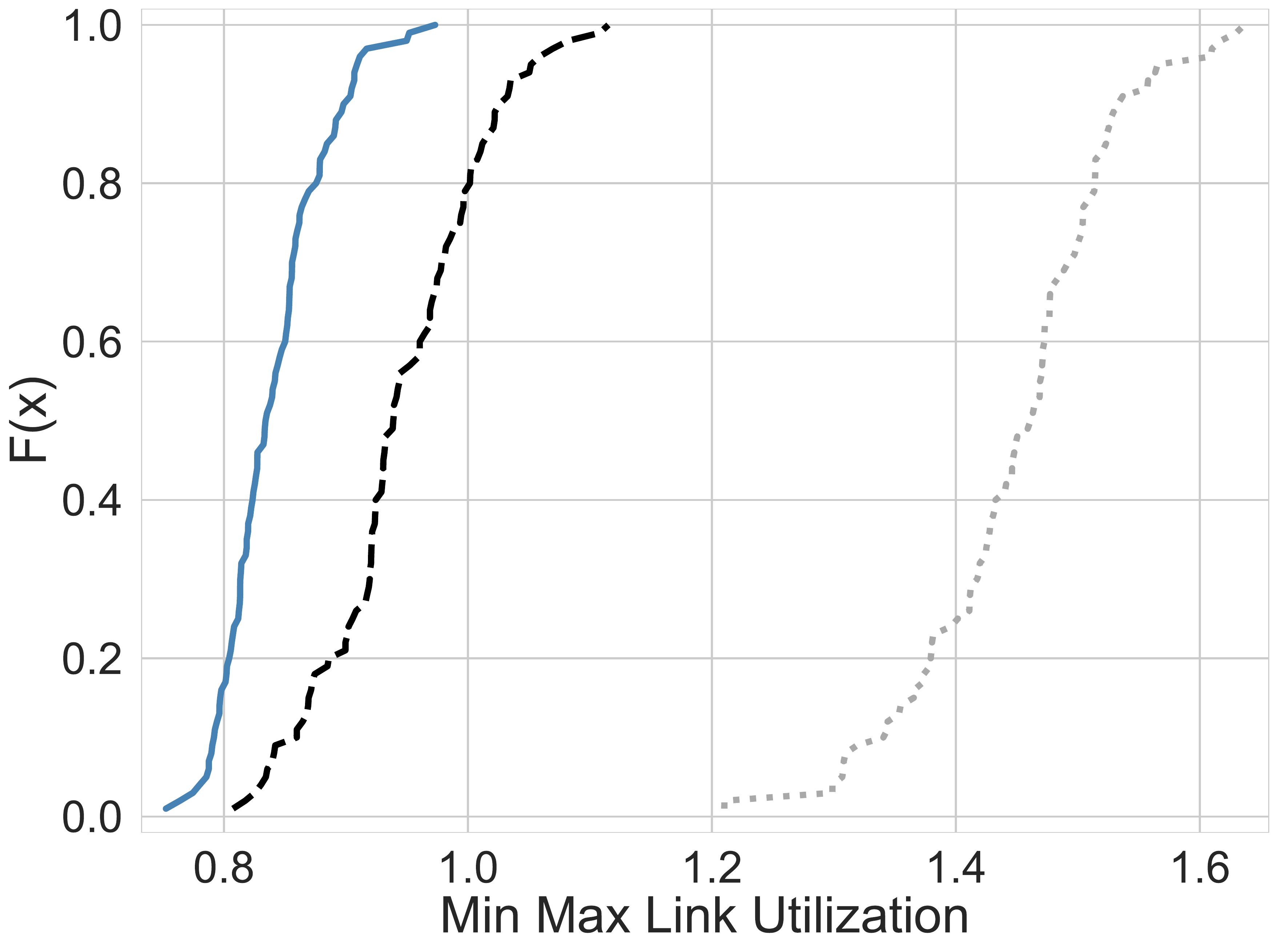}
        \caption{CDF GEANT2}
	\label{subfig:CDF_GEANT2_uniform}
    \end{subfigure}
     \caption{Evaluation results of Minimum Maximum Link Utilization with uniform traffic profiles in the NSFNet and GEANT2 network topologies. The evaluation is done over 100 traffic matrices unseen during training.}
  \label{fig:tm_uniform}
     \vspace{-0.2cm}
\end{figure}

Figure \ref{fig:tm_uniform} shows the evaluation result considering a uniform traffic profile. For the sake of readability, these plots show both the raw Minimum Maximum Link Utilization values obtained for each TM, and the Cumulative Distribution Function (CDF) of these results. In this case, we can observe that our proposed MARL+GNN solution performs significantly better than default OSPF in both topologies (on average, $\approx$23\% better in NSFNet and $42\%$ in GEANT2) and stays near to the close-to-optimal solutions produced by DEFO algorithm (in GEANT2, it even improves it by $11\%$).

\begin{figure}[!h]
    \begin{subfigure}[]{0.495\columnwidth}
	\includegraphics[width=1.0\linewidth,height=3.5cm]{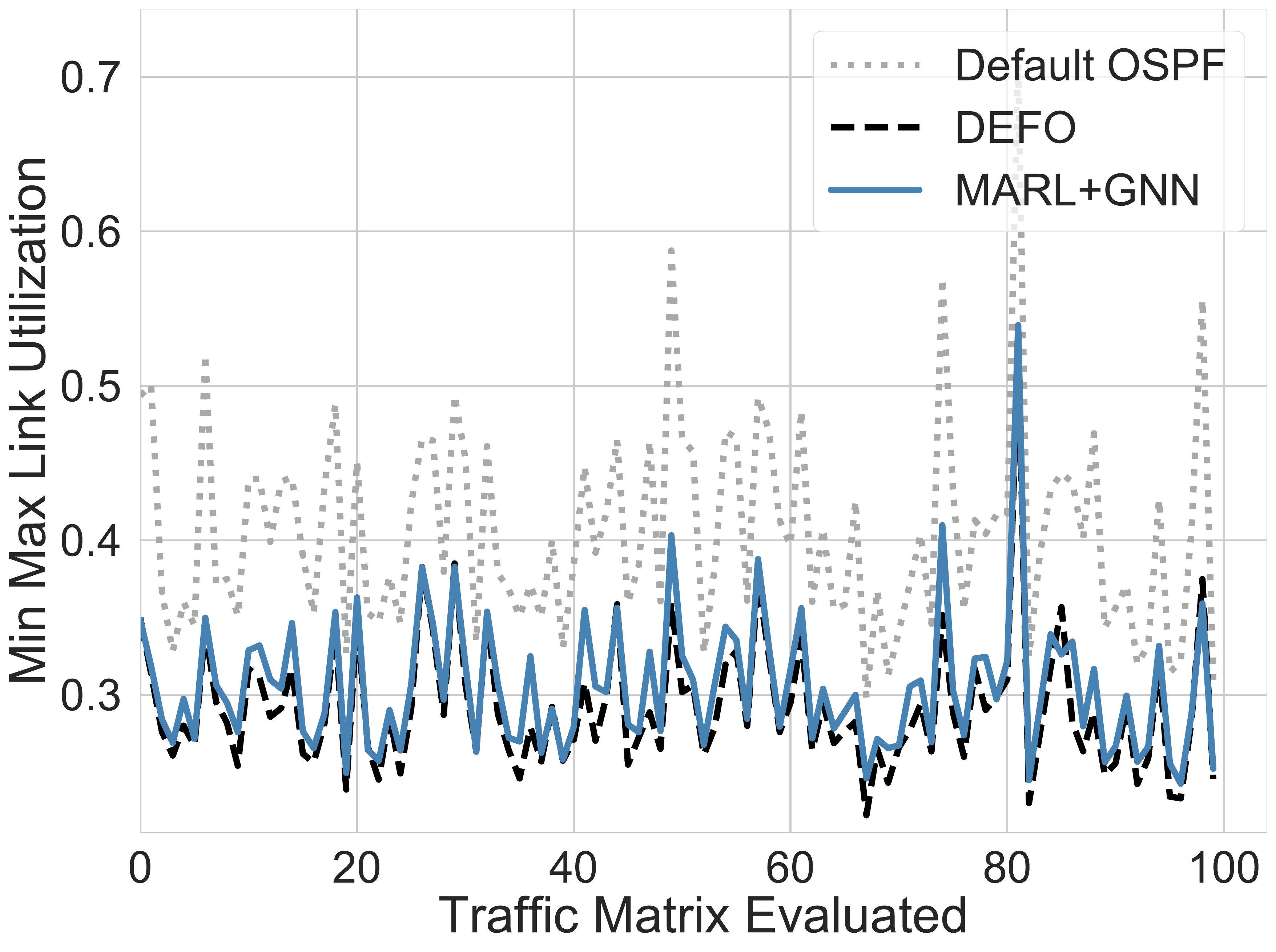}
        \caption{\textit{MinMaxLoad} NSFNet}
	\label{subfig:raw_NSFNet_gravity}
    \end{subfigure}
    \begin{subfigure}[]{0.495\columnwidth}
	\includegraphics[width=1.0\linewidth,height=3.5cm]{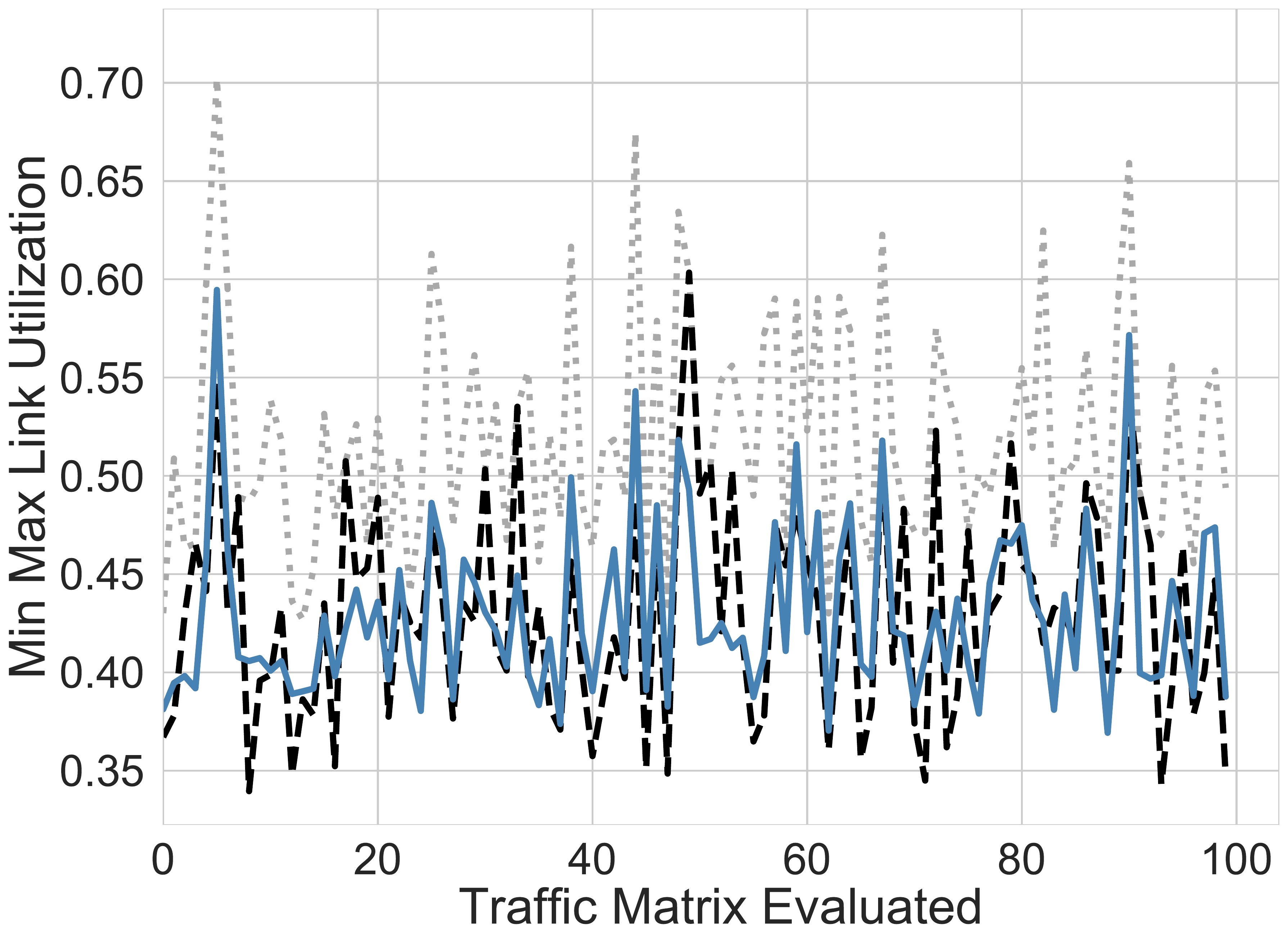}
        \caption{\textit{MinMaxLoad} GEANT2}
	\label{subfig:raw_GEANT2_gravity}
    \end{subfigure}
    \begin{subfigure}[]{0.495\columnwidth}
	\includegraphics[width=1.0\linewidth,height=3.5cm]{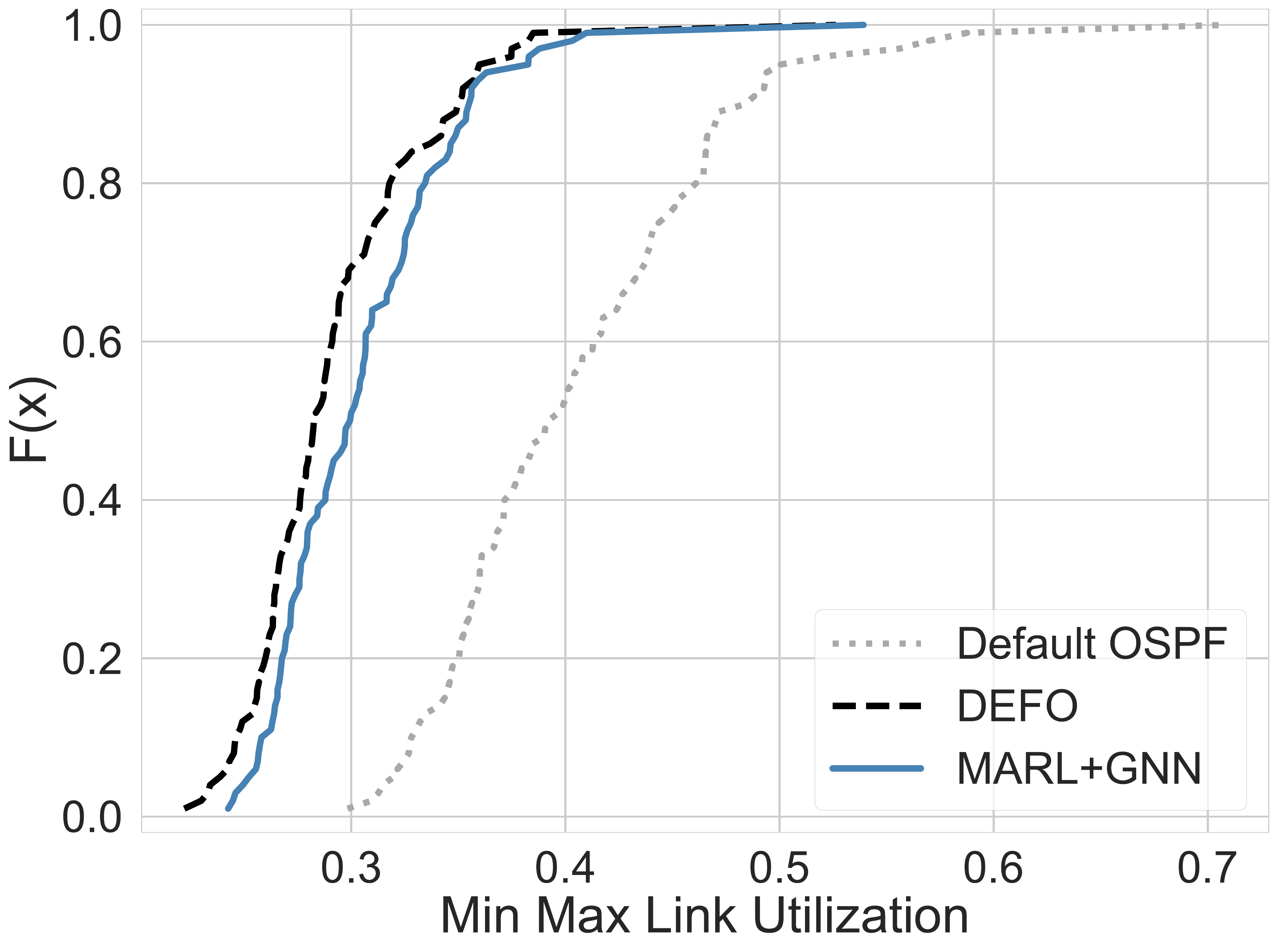}
        \caption{CDF NSFNet}
	\label{subfig:CDF_NSFNet_gravity}
    \end{subfigure}
    \begin{subfigure}[]{0.495\columnwidth}
	\includegraphics[width=1.0\linewidth,height=3.5cm]{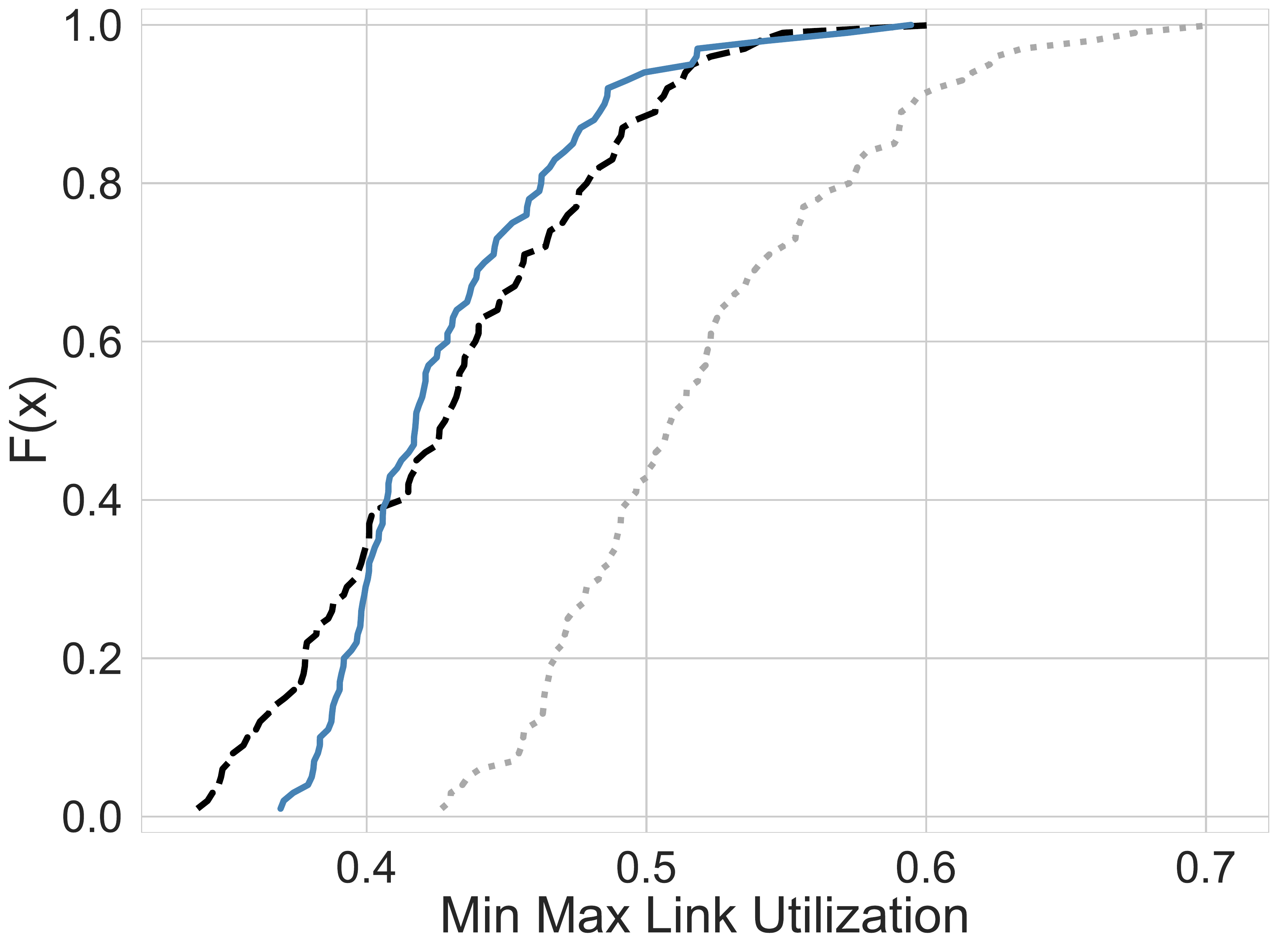}
        \caption{CDF GEANT2}
	\label{subfig:CDF_GEANT2_gravity}
    \end{subfigure}
     \caption{Evaluation results of Minimum Maximum Link Utilization with gravity-based traffic profiles in the NSFNet and GEANT2 network topologies. The evaluation is done over 100 traffic matrices unseen during training.}
  \label{fig:tm_gravity}
     \vspace{-0.4cm}
\end{figure}

Analogously, Figure \ref{fig:tm_gravity} presents the evaluation results in scenarios with the gravity traffic profile. Again, our proposed MARL+GNN solution outperforms default OSPF in both topologies (on average, $\sim$25\% better in NSFNet and $17\%$ in GEANT2) and attains a comparable performance to DEFO. 

\subsection{Generalization over other Network Topologies} \label{subsec:res-topology}

While traditional TE optimizers are typically designed to operate on arbitrary networks, current state-of-the-art ML-based solutions for TE suffer from a lack of topology generalization, partly explained by the fixed-size input scheme of most ML models (fully-connected NNs, convolutional NNs). That is, previous ML solutions could only operate on those toplogies seen during the training phase. Therefore, achieving generalization over different topologies is an essential step towards the versatility of state-of-the-art classical TE methods. 

Given that our distributed GNN-based proposal naturally allows variable-size network scenarios, as well as relational reasoning~\cite{battaglia2016interaction,suarez2019challenging}, we are particularly interested in evaluating the generalization potential of our MARL+GNN solution over other networks not considered in training. For these experiments, we train our model in both NSFNet and GEANT2 topologies, and then evaluate it in a never-seen network (GBN). In this case, we stop the training when the system observes a total of $100$ TMs --alternating NSFNet and GEANT2 instances every $50$ training episodes-- and evaluate it over $100$ TMs in GBN. 

\begin{figure}[!t]
\centering
	\begin{subfigure}[]{0.493\columnwidth}
	\includegraphics[width=1.0\linewidth,height=3.4cm]{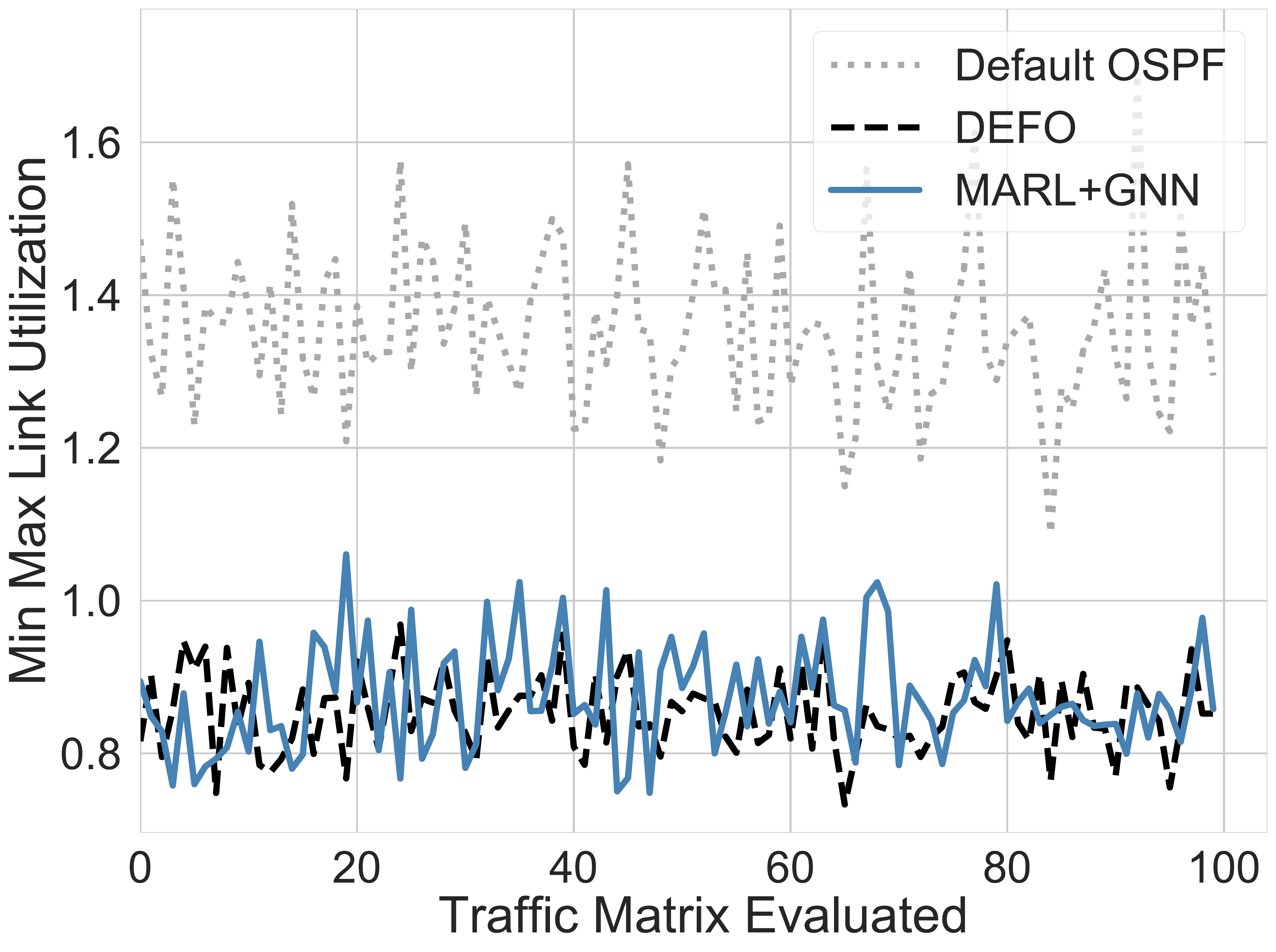}
        \caption{\textit{MinMaxLoad} GBN}
	\label{subfig:rax_GBN}
    \end{subfigure}
    \begin{subfigure}[]{0.493\columnwidth}
	\includegraphics[width=1.0\linewidth,height=3.4cm]{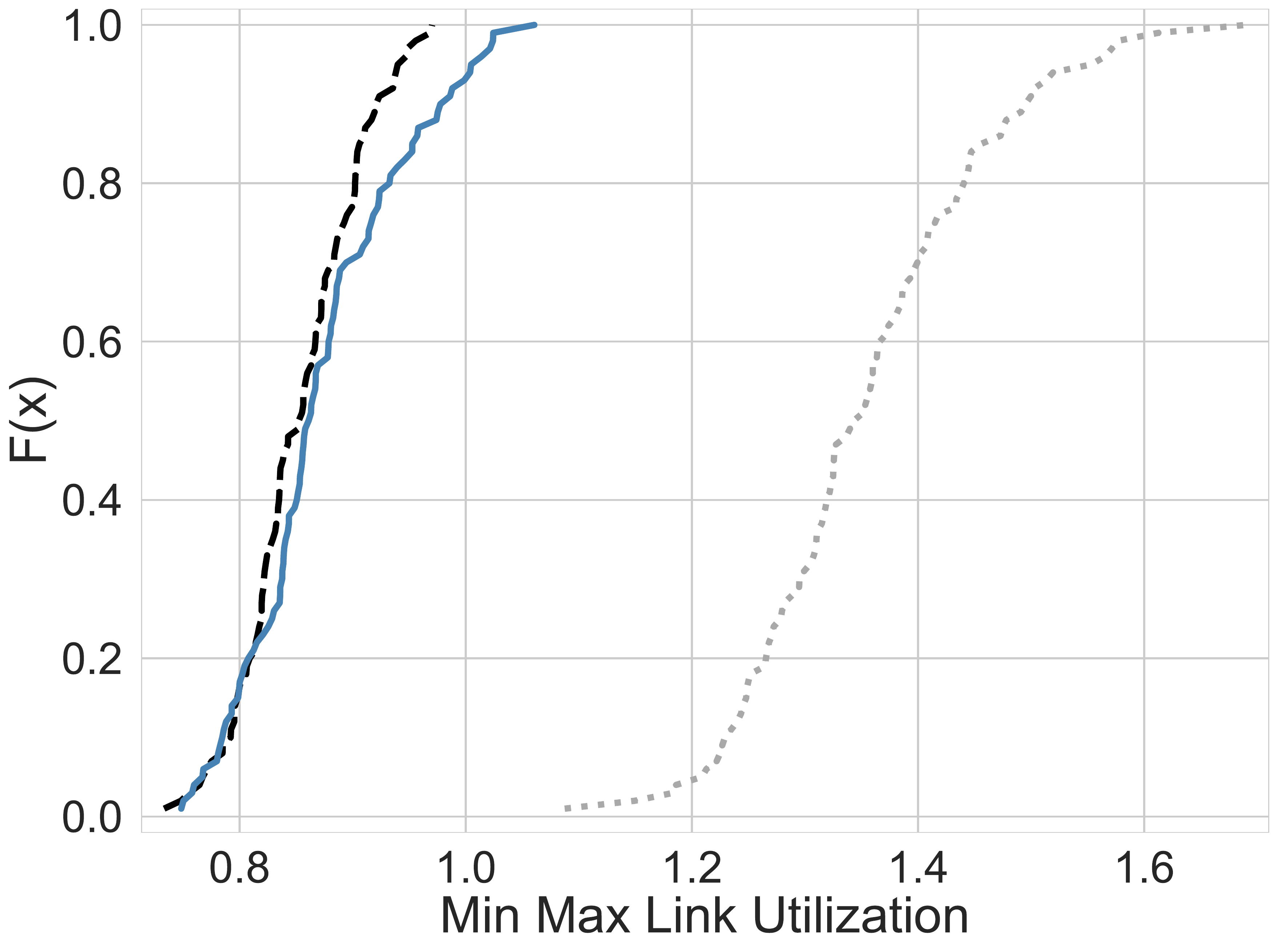}
        \caption{CDF GBN}
	\label{subfig:CDF_GBN}
    \end{subfigure}
  \caption{Evaluation results of Minimum Maximum Link Utilization for 100 different configurations in GBN after training the model using exclusively with samples of NSFNet and GEANT2.}
  \label{fig:cdf_datanet}
  \vspace*{-0.2cm}
\end{figure}

Figure \ref{fig:cdf_datanet} presents the evaluation results of this experiment, showing the Minimum Maximum Link Utilization values obtained at each sample, as well as the CDF of these results. Here we can observe that the proposed solution significantly outperforms default OSPF ($35\%$ better on average) and it is very close -only within a $2\%$ difference- to DEFO.

\subsection{Robustness against Link Failures}

The ability to generalize over different network topologies opens the door to address other uses cases that could not be solved with previous ML-based solutions. For example, in this section we assess how our solution performs when the network experiences link failures, which inevitably result in changes in the topology.
To this end, we design the following experiment: given a traffic matrix and a topology, our model previously trained in Section~\ref{subsec:res-topology} is applied in networks with increasing number of random link failures \mbox{--up} to a maximum of $9$ failures. We repeat this experiment $10$ times for a given number of failures $n$, exploring at each iteration different combinations of link failures.

\begin{figure}[!t]
\centering
\includegraphics[width=\columnwidth, height=0.4\columnwidth]{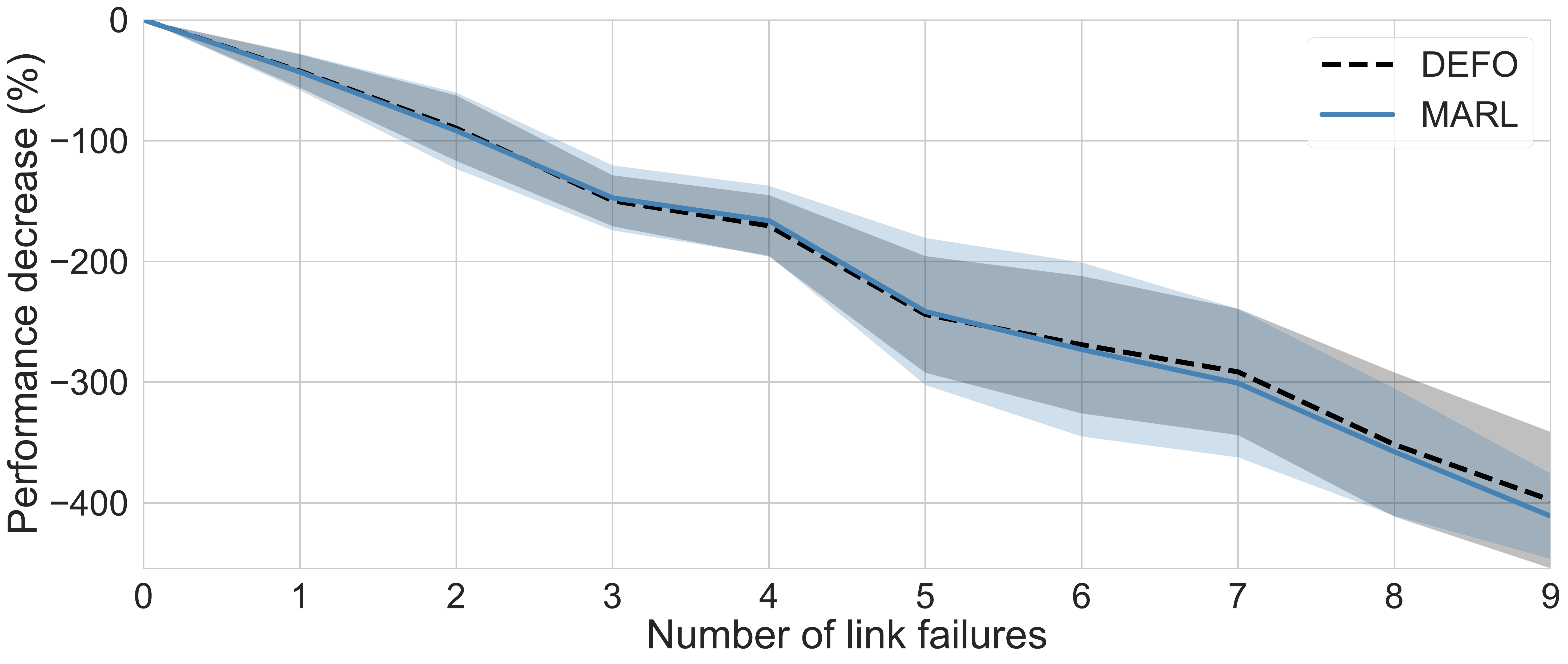}
  \caption{Performance degradation with increasing link failures for our NSFNet+GEANT2 model (applied to GBN), and DEFO. The plot shows the mean and standard deviation for 5 different TMs; for each TM we average the results on 10 scenarios with \textit{n} random link failures.}
  \label{fig:link_failure}
  \vspace*{-0.2cm}
\end{figure}

Figure~\ref{fig:link_failure} shows the mean and standard deviation of the performance degradation --w.r.t. the original network scenario with all the links-- over $5$ different traffic matrices, using the model trained exclusively in NSFNet and GEANT2 (Sec.~\ref{subsec:res-topology}) and applying it over the GBN network topology. These results are compared against DEFO, which is evaluated under the same conditions (i.e., same TMs and network scenarios). As we can observe, the performance decays gracefully as the number of removed links increases, showing an almost identical behavior to that of the state-of-the-art DEFO technique.

\begin{table*}[ht]
\centering
\begin{tabular}{@{}lccccc@{}}
\toprule
                             & \textbf{NSFNet}  &   \textbf{GBN}   & \textbf{GEANT2}  &  \textbf{SYNTH500}    & \textbf{SYNTH1000}    \\ \midrule
Episode Length               &     100          &    150           &      200         &       5,250       &     9,600          \\
Execution Time (s)                     &$9.98\cdot10^{-2}$&$1.33\cdot10^{-1}$&$2.12\cdot10^{-1}$&       8.40       &     19.2          \\
Average MPNN-based Link Overhead$^{*}$ (MB/s) &     1.20         &     1.32         &     1.20         &       1.60       &     1.41          \\
 \bottomrule
\multicolumn{6}{l}{$^{*}$It includes a $20\%$ extra cost per message considering headers and metadata.}
\end{tabular}
\caption{Cost of our solution -- Execution time and average link overhead. Applied to variable-sized network topologies, and assuming that hidden states are encoded as $16$-element vectors of floats, and each Message Passing runs K=$8$ steps.}
\label{tab:MP_cost}
\end{table*}

\subsection{Performance vs. Message Passing Iterations} \label{subsec:steps}

Previous experiments have shown that the proposed solution achieves comparable performance to DEFO across a wide variety of scenarios. However, there are still another important feature: \textit{the execution cost}, which can be a crucial aspect to assess whether the proposed ML-based solution can achieve reasonable execution times for near real-time operation, as in DEFO.

With the above objective in mind, in this section we first analyze the impact of one main hyperparameter of our MARL system, which is the episode length $T$. This is the maximum number of optimization steps that the MARL system needs to execute before producing a good set of link weights. Given that in our framework only one of the agents selects an action (i.e., increase its link weight) at each time-step of the episode, we expect a straightforward correlation between the number of steps and the total amount of links in the network: the larger the number of links, the larger should potentially be the episode explorations to achieve a good configuration. Finding the exact relation, though, depends on multiple complex factors (e.g., the distribution of links, the initialization of weights, the estimated traffic demands).

\begin{figure}[!t]
\centering
	\includegraphics[width=\columnwidth, height=0.4\columnwidth]{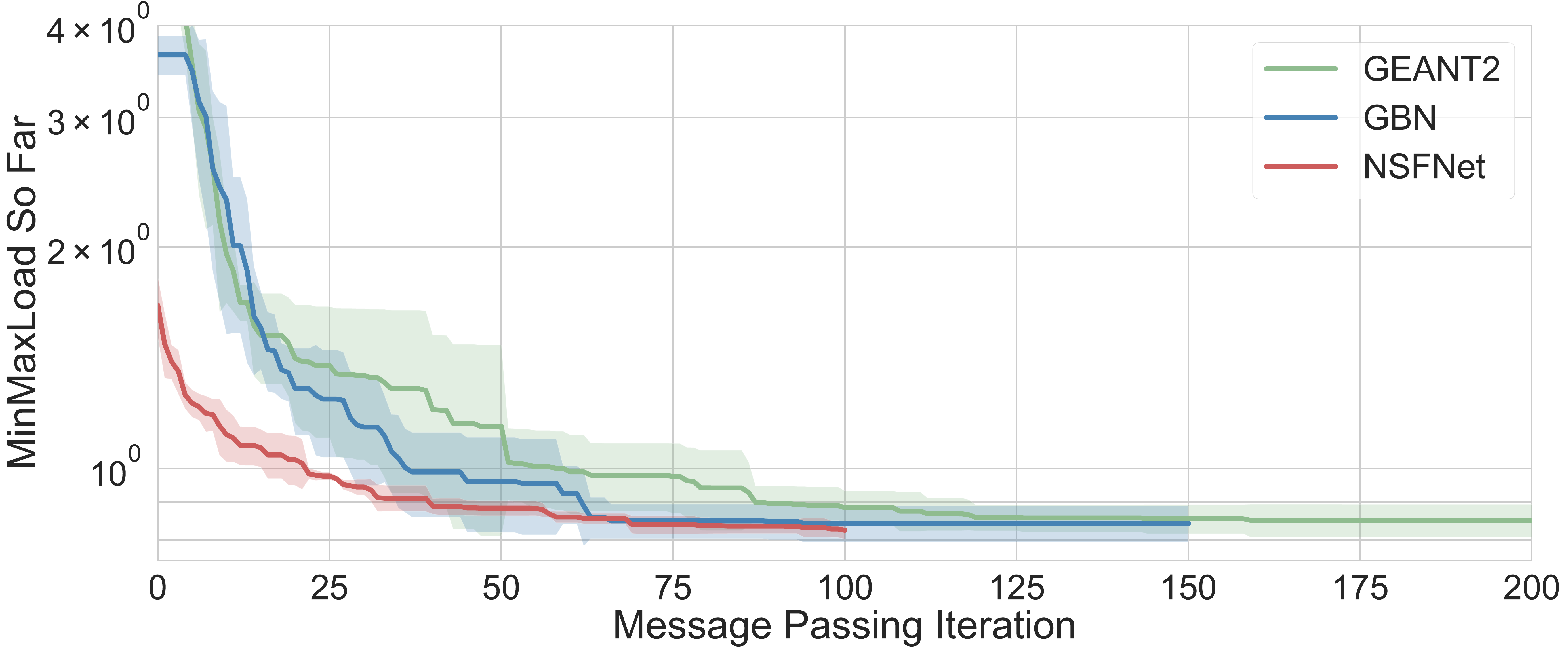}
  \caption{Evolution of the MinMax link load so far along an episode when applying our model (trained in NSFNet+GEANT2) respectively to NSFNet, GBN, and GEANT2. Plots show the mean and std. dev. over 5 runs, each considering a different TM.}
  \label{fig:message_iterations}
  \vspace*{-0.2cm}
\end{figure}

By exploring systematically a variable number of steps in the three topologies considered above (NSFNet, GBN, GEANT2), we have empirically found that with an episode length $\approx$2-3 times the number of links in the network, our system reaches its best performance --which is comparable to the near-optimal results of DEFO, as observed in previous sections. For instance, in our experiments it is sufficient to define $T$=100 for NSFNet, $T$=150 for GBN, and $T$=200 for GEANT2. This can be observed in Figure~\ref{fig:message_iterations}, which shows the evolution of the maximum link utilization achieved by our MARL system along an episode in the three network topologies.

\subsection{Cost Evaluation} \label{subsec:res-cost}

Considering the previous evaluation on the episode length (Sec.~\ref{subsec:steps}), in this section we aim to evaluate the execution time of our solution to reach its best optimization potential. This is probably the main advantage that we can expect from ML-based solutions w.r.t. to near-optimal state-of-the-art TE techniques, such as DEFO. Indeed, if we analyze the main breakthroughs of DRL in other fields (e.g., \cite{schrittwieser2020mastering}), we can observe they have been mainly achieved in complex online decision-making and automated control problems. Note also that, after training, our multi-agent system is deployed in a distributed way over the network, thus distributing the computation of the global TE optimization process. 

Table \ref{tab:MP_cost} shows the execution time of our MARL+GNN trained system for the three real-world topologies used in our evaluation: NSFNet, GBN and GEANT2. Moreover, we also simulated executions over two synthetic networks \mbox{--SYNTH500} (500 nodes, 1750 links) and SYNTH1000 (1000 nodes, 3200 \mbox{links)--} in order to analyze the cost of our distributed system in larger networks. As we can see, the execution time of our solution scales in a very cost-effective way with respect to the size of the network; from the order of milliseconds in NSFNet, GBN and GEANT2, to the tens of seconds in the SYNTH1000 network, with thousands of nodes and links. In contrast, DEFO requires 3 minutes for optimizing networks of several hundreds of nodes~\cite{hartert2015declarative}. This shows an important reduction in the execution cost of our solution; particularly it represents a one-order-of-magnitude improvement in the case of the largest network (SYNTH1000).

We note, though, that this improvement is achieved at the expense of exchanging additional GNN messages between nodes (MPNN). We show in Table \ref{tab:MP_cost} the MPNN communication cost in terms of the average link overhead resulting from such extra messages. As expected, the cost is quite similar in all topologies, as the messaging overhead of our distributed protocol is directly proportional to the average node degree (i.e., number of neighbors) of the network, and computations are distributed among all nodes. In particular, we can see that the average link overhead only involves a bandwidth of few MB/s per link independently of its capacity, which can reasonably have a negligible impact in today's real-world networks with 10G/40G (or even more) interfaces.

\section{Related Work} \label{sec:related}

Network optimization is a well-known and established topic whose fundamental goal is to operate networks efficiently. Most of the work in the literature uses classical methods to optimize the network state (e.g., ILP). During the last decade a plethora of algorithms have been proposed, exploring a wide spectrum of techniques and network \mbox{abstractions~\cite{hartert2015declarative,gay2017expect,bhatia2015optimized,jadin2019cg4sr}.}

Some works have previously attempted to apply DRL~\cite{valadarsky2017learning,xu2018experience,almasandeep} or MARL~\cite{ding2020packet,geng2020multi} to TE. However, they were unable to report a performance comparable to the state of the art, as they were compared to simpler routing schemes, such as SP routing (e.g., \cite{xu2018experience,ding2020packet,geyer2018learning}), SP+ECMP (e.g., \cite{geng2020multi}), Load Balancing (e.g., \cite{xu2018experience,almasandeep}) or oblivious routing (e.g., \cite{valadarsky2017learning}). Our work is the first to be benchmarked against a state-of-the-art optimizer --i.e., DEFO~\cite{hartert2015declarative}-- and to provide enough evidence to address the open question posed in this paper. 

Moreover, most of the current state-of-the-art (MA)RL-based TE solutions \cite{valadarsky2017learning,xu2018experience,ding2020packet} suffer from another limitation: they fail to generalize to unseen scenarios (e.g., different network topologies) as the implemented traditional neural networks (e.g., fully connected, convolutional) are not well-suited to learn and generalize over data that is inherently structured as graphs. One exception is the work of \cite{geng2020multi}, a different MARL approach that addresses an inter-region TE network scenario. 

GNNs~\cite{gori2005new, scarselli2008graph}, and in particular MPNNs \cite{gilmer2017neural}, precisely emerged as specialized methods for dealing with graph-structured data; for the first time, there was an AI-based technology able to provide with topology-aware systems. In fact, GNNs have recently attracted a larger interest in the computer networks field for addressing the aforementioned generalization limitations. The work from \cite{rusek2019unveiling} proposed the use of GNNs to predict network performance metrics (e.g., average delay). Authors of \cite{almasandeep} proposed a novel architecture for routing optimization in Optical Transport Networks that embeds a GNN into a centralized, single-agent RL setting that is compared against Load Balancing routing.

Arguably, the closest work to this paper is \cite{geyer2018learning}, whose premise is similar to ours: the generation of easily-scalable, automated distributed protocols for intradomain TE. For doing so, the authors of this work also make use of a GNN, but unlike our unsupervised RL-based solution, their approach is limited to learn already existing protocols (e.g., shortest path routing). Particularly, they use a semi-supervised learning algorithm that requires labeled data, against which they compare their solution. In fact, there are very few works that have combined GNN with a DRL framework~\cite{jiang2020graph, su2020counterfactual}, and they are theoretical proposals from the ML community that do not apply to the field of networking.

\section{Discussion \& Concluding Remarks} \label{sec:conclusions}

This paper started asking an open question: \textit{Is ML ready for Traffic Engineering Optimization?} To answer it, we proposed a novel distributed GNN-based MARL architecture --which represents the state-of-the-art in ML-- and compared it against the state-of-the-art in TE. From our qualitative and experimental analysis, we derive the following conclusions:

\textbf{Performance:} ML can attain comparable performance to the state-of-the-art in TE over unseen traffic and topologies. While previous ML-based solutions for TE (see Section \ref{sec:related}) have shown significant improvements over basic routing methods --such as shortest path, SP+ECMP or load balancing-- they have not been compared yet to advanced TE solutions \mbox{--like} \mbox{DEFO~\cite{hartert2015declarative}--}. We present for the first time a ML-based system that obtains similar performance --and even better in some scenarios-- to state-of-the-art TE optimizers.

\textbf{Speed and Hardware:} ML can obtain close-to-optimal results much faster than the state-of-the-art in TE. While our MARL+GNN solution can optimize a network in the scale of seconds, DEFO operates in the range of few minutes. This is mainly due to the fact that our distributed architecture naturally parallelizes the global optimization process among all network devices (i.e., routers). In contrast, DEFO and other advanced TE optimizers~\cite{gay2017expect,bhatia2015optimized,jadin2019cg4sr} use a centralized approach that cannot benefit from this. Moreover, since network devices would need to run a GNN, our proposal could also profit from the use of specialized AI chips. Substantial research efforts are currently being devoted into designing custom GNN hardware accelerators \cite{DBLP:journals/corr/abs-2001-02498} that aim for 10x speed improvements over current GPUs.

\textbf{Refined Goals:} Our results suggests that current ML technologies are becoming mature enough to deal with complex networking scenarios. However, we acknowledge that AI-based methods still lack the versatility of traditional TE optimizers to face very different scenarios and/or several objectives at the same time. As an example, state-of-the-art TE optimizers can operate at the ingress-egress or even at flow granularities, producing TE configurations that can optimize the network resources while fulfilling SLA requirements for specific connections. Our ongoing work is focused on the design of objective-agnostic MARL agents that can handle multiple, arbitrary TE goals using meta-learning methods.

\bibliographystyle{ieeetr} 
\bibliography{References.bib}

\end{document}